\documentclass[aps,twocolumn,prb,superscript,floatfix,superscriptaddress,showpacs]{revtex4-1}
\usepackage{amssymb}

\usepackage{epsf}
\usepackage{epsfig}
\usepackage{graphicx}
\usepackage{dcolumn}
\usepackage{braket}
\usepackage{bm}
\usepackage{amsfonts}
\usepackage{amsmath}
\usepackage{amssymb}
\usepackage{color,soul}
\usepackage{wasysym}
\usepackage{mathrsfs}
\usepackage{natbib}
\usepackage{multirow}
\usepackage[colorlinks=true,%
            linkcolor=blue,%
            urlcolor=blue,%
            citecolor=blue,%
            filecolor=blue,%
            bookmarksopen=true,%
            pdfauthor={J. P. Ramos Andrade},%
            pdftitle={THERMO_MBS},%
            pdfsubject={THERMO_MBS_Manuscript},%
            pdfpagemode=UseOutlines]{hyperref}

\newcommand{\ve}{\varepsilon}

\newcommand{\bea}{\begin{eqnarray}}
\newcommand{\eea}{\end{eqnarray}}

\newcommand{\la}{\langle}
\newcommand{\ra}{\rangle}

\begin{document}

\title{Thermoelectric transport through Majorana bound states and violation of Wiedemann-Franz law}
\author{J. P. Ramos-Andrade}
\email{juan.ramosa@usm.cl}	
\affiliation{Departamento de F\'isica, Universidad T\'ecnica Federico Santa Mar\'ia, Casilla 110 V, Valpara\'iso, Chile}
\affiliation{Department of Physics and Astronomy, and Nanoscale and Quantum Phenomena Institute, Ohio University, Athens, Ohio 45701-–2979, USA}
\author{O. \'Avalos-Ovando}
\affiliation{Department of Physics and Astronomy, and Nanoscale and Quantum Phenomena Institute, Ohio University, Athens, Ohio 45701-–2979, USA}
\author{P. A. Orellana}
\affiliation{Departamento de F\'isica, Universidad T\'ecnica Federico Santa Mar\'ia, Casilla 110 V, Valpara\'iso, Chile}
\author{S. E. Ulloa}
\affiliation{Department of Physics and Astronomy, and Nanoscale and Quantum Phenomena Institute, Ohio University, Athens, Ohio 45701-–2979, USA}

\begin{abstract}

We study features of the thermoelectric transport through a Kitaev chain hosting Majorana bound states (MBS) at its ends. We describe the behavior of the Seebeck coefficient and the $ZT$ figure of merit for two different configurations between MBS and normal current leads. We find an important violation of the Wiedemann-Franz law in one of these geometries, leading to sizeable values of the thermoelectric efficiency over a narrow window in chemical potential away from neutrality. These findings could lead to interesting thermoelectric-based MBSs detection devices, via measurements of the Seebeck coefficient and figure of merit.

\end{abstract}
\pacs{}
\date{\today}
\maketitle

\section{Introduction}

A new kind of fermionic quasi-particle has been studied in the context of condensed matter in recent years, with its principal feature being that it is its own antiparticle. These Majorana fermions (MFs), first predicted by E. Majorana,\cite{majorana1937teoria} have other interesting properties such as satisfying non-Abelian statistics and are therefore of interest in quantum computation implementations.\cite{alicea2016superconductors,alicea2011non} These quasi-particles appear in systems with particle-hole symmetry as zero-energy excitations, and are predicted to be found at the ends of a one-dimensional semiconductor nanowire with spin orbit interaction (SOI) in a magnetic field and proximitized by an adjacent superconductor.\cite{Roman2010Majorana,Yuval2010helical} Such Majorana states may also appear in other systems as in a vortex of a $p$-wave superconductor,\cite{Ivanov2001non} on the surface of a topological insulator,\cite{Fu2008superconducting} and at the ends of a chain of magnetic impurities on a superconducting surface.\cite{nadj2014observation,ruby2015end} The Majorana bound states (MBS) at the end of such a wire/chain system, can be seen as implementation of a Kitaev chain.\cite{kitaev2001unpaired} Mourik \emph{et al.}\cite{mourik2012signatures} reported the first observation of Majorana signatures in a semiconductor-superconductor nanowire, built of InSb (indium antimonide) and  NbTiN (niobium titanium nitride), with several others groups reporting zero-bias conductance peaks in similar hybrid devices.\cite{deng2012anomalous,das2012zero,Churchill2013superconductor,rokhinson2012fractional} MBS pairs are predicted to interact with a coupling strength $\ve_{M}$ proportional to $\exp[-\mathcal{L}/\xi]$, where $\mathcal{L}$ is the wire length and $\xi$ is the superconducting coherence length. Recent experimental work has probed this dependence of $\ve_{M}$ in wire length, verifying expectation.\cite{albrecht2016exponential}\\

Moreover, there is a great deal of interest in the thermoelectricity of nanostructures.\cite{diventra2011thermoelectricity,bauer2012Spincaloritronics,sanchez2014} When a thermal bias is applied across a system, a quantity of interest is the thermoelectric energy-conversion efficiency, characterized by the dimensionless figure of merit $ZT$, which involves the Seebeck coefficient, as well as the ratio of thermal and electrical conductances.\cite{goldsmid2013thermoelectric} A way to improve $ZT$ is to overcome the Wiedemann-Franz law, which sets the ratio $\kappa/\mathcal{G}T=\ell_{0}\equiv\text{constant}$ in all systems, where $\mathcal{G}$ is the electrical conductance, $\kappa$ the thermal conductance, $T$ the background temperature and $\ell_{0}=(\pi^{3}/3)(k_{B}/e)^{2}$ is the Lorenz number.\cite{franz1853ueber} Although macroscopic materials have shown to generally follow the Wiedemann-Franz law, nanostructured systems have proved to be very good thermoconverters as they are able to overcome that restriction.\cite{Vineis2010NanostructuredThermoelectrics} Thermoelectric efficient devices have been proposed in systems such as molecular junctions,\cite{bergfield2009thermoelectric,Reddy2007Thermoelectricity} quantum dots\cite{Trocha2012LargeEnhancement} and topological insulators.\cite{Tretiakov2011Holey} Thermal detection of Majorana states in topological superconductors has also been proposed.\cite{beenakker} Even though several Majorana detection setups have been realized,\cite{mourik2012signatures,deng2012anomalous,das2012zero,Churchill2013superconductor,rokhinson2012fractional,nadj2014observation} much less attention has been directed to thermoelectric-based detection devices. Different thermoelectric-setups with Majorana nanowires and/or connected quantum dots have been considered, where thermal biases are applied across the normal leads\cite{lopez2014thermoelectrical} or across normal lead-superconductor setups.\cite{leijnse2014} These systems are found to exhibit signatures of MBS through measurements of the Seebeck coefficient as the energy of the level in the dot varies, even in a weak coupling regime.\\


In this work we study the thermoelectrical properties of a MBS system coupled to two normal leads in the presence of a thermal bias. We model the system as a Kitaev chain hosting two MBSs, $\gamma_{1}$ and $\gamma_{2}$, coupled between them with a strength $\ve_{M}$ (assumed known). Using a Green's function formalism, we study the thermoelectric transport across the Kitaev chain, in two different configurations: i) when both MBSs are connected to the leads, and ii) when only one MBS is connected to the leads. The first configuration was discussed on Ref.\ [\onlinecite{lopez2014thermoelectrical}] for the case of zero chemical potential ($\mu=0$) in contacts. Our findings agree with their results and go further as chemical potential varies. We find a small Seebeck coefficient and vanishing small $ZT$ over broad range of chemical potential and coupling $\ve_{M}$ at typical low experiment temperatures. On the other hand, we find a sizeable violation of the Wiedemann-Franz law for the second configuration, which leads to large values of thermoelectric efficiency, as measured by the figure of merit. We also find an $\ve_{M}$-independent behavior of the thermal quantities with  large $\ve_{M}$ values for the same configuration. These features should be accessible in experiments and may help provide additional insights into the presence of behavior of MBSs in nanowire systems.\\

This paper is arranged as follows: Section\ \ref{secmodel}, presents the model and Hamiltonian used for obtaining the thermoelectric quantities. Section\ \ref{results} shows the results and discussion; and finally the concluding remarks are in Section\ \ref{secconclusions}.

\section{Model}\label{secmodel}

\begin{figure}[tbph]
\centering
\includegraphics[width=0.48\textwidth]{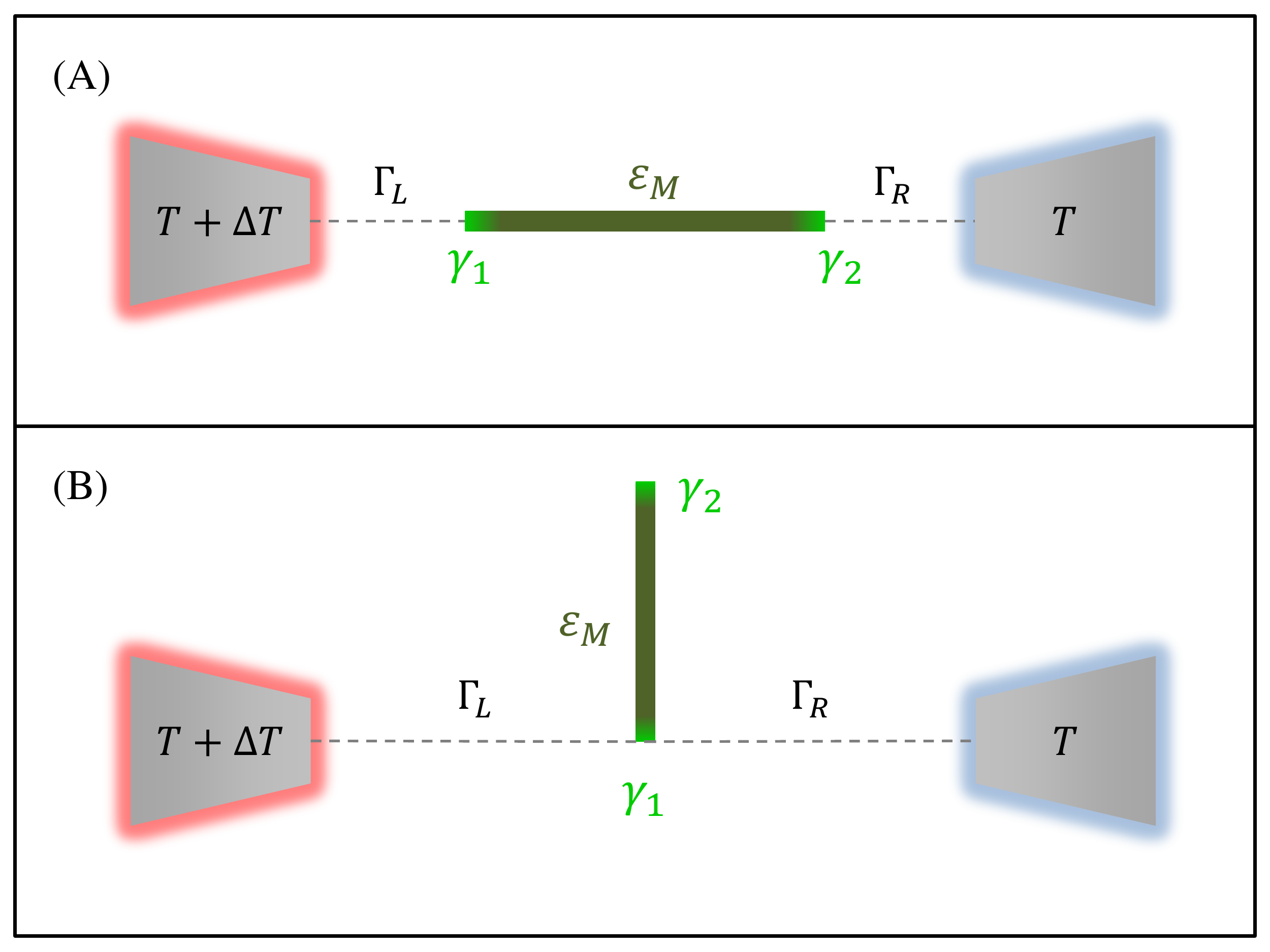}
\caption{(Color online) Model setup of a Kitaev chain hosting MBS at both ends, connected to two metallic leads at different temperatures. In (A) each MBS is coupled with its nearest lead while in (B) only one MBS is coupled to both leads simultaneously.}
\label{Model}
\end{figure}

We consider a two-MBS system, each located at the ends of a Kitaev chain and coupled to two metallic leads in two different configurations, as shown schematically in Fig.\ \ref{Model}. The left lead $L$ is kept at temperature $T+\Delta T$ and the right lead $R$ at temperature $T$, providing thus a temperature gradient $\Delta T$. We describe the system with a noninteracting Anderson Hamiltonian within the second quantization framework, and consider it as spin-independent because of a strong Zeeman effect due to the applied magnetic field. The Hamiltonian is given by\cite{lopez2014thermoelectrical}
\begin{equation}
H=H_{\text{leads}}+H_{\text{leads-MBS}}+H_{\text{MBS}}\,,\label{Hamiltonian}
\end{equation}
where $H_{\text{leads}}$ describes the current leads, $H_{\text{leads-MBS}}$ the coupling between leads and MBS, and $H_{\text{MBS}}$ the isolated MBSs. Each of them is given by
\begin{eqnarray}
H_{\text{leads}}&=&\sum_{\alpha,k}\ve_{\alpha,k}c_{\alpha,k}^{\dag}c_{\alpha,k}\,, \label{Hleads} \\
H_{\text{MBS}}&=&i\ve_{M}\gamma_{1}\gamma_{2}\,, \label{HMBS} \\
H_{\text{leads-MBS}}&=&\sum_{\alpha,k}t_{\alpha,\beta}\gamma_{\beta}c_{\alpha,k}+t_{\alpha,\beta}^{\ast}c_{\alpha,k}^{\dag}\gamma_{\beta}\,, \label{HleadsMBS}
\end{eqnarray}
where $c_{\alpha,k}^{\dag} (c_{\alpha,k})$ creates (annihilates) an electron of momentum $k$ in lead $\alpha=L,R$, $\gamma_{\beta}$ creates one of the two MBS ($\beta=1,2$), and satisfies both $\{\gamma_{\beta},\gamma_{\beta'}\}=2\delta_{\beta,\beta'}$ and $\gamma_{\beta}=\gamma_{\beta}^{\dag}$, \emph{i.e.} a MBS is its own antiparticle. $\ve_{M}$ is the coupling between the two MBSs due to a finite length of the wire. The terms $t_{\alpha,\beta}$ are the tunneling hoppings between the lead $\alpha$ and the MBS $\beta$. For the two models shown in Fig.\ \ref{Model}, the upper and lower panels consider $t_{L,1}=t_{R,2}\neq 0$ and $t_{L,1}=t_{R,1}\neq 0$, respectively, with others vanishing.\\

We obtain the transmission probability across the leads, by using the Green's function formalism. In the linear response regime, we can obtain the transmission by means of the Fischer-Lee relation, given by
\begin{equation}
\mathcal{T}(\ve)=\text{Tr}\left[\tilde{\Gamma}_{L}\tilde{G}^{a}(\ve)\tilde{\Gamma}_{R}\tilde{G}^{r}(\ve)\right]\,, \label{Tras}
\end{equation}
with $\ve$ the energy of the electron tunneling from $L$ to $R$, $\tilde{\Gamma}_{\alpha}$ being the coupling matrix of the lead $\alpha$ and $\tilde{G}^{r}(\ve)$ ($\tilde{G}^{a}(\ve)$) the retarded (advanced) Green's function matrix given by
\begin{equation}
\tilde{G}^{r}(\ve)=\left(
                     \begin{array}{cc}
                       \la\la \gamma_{1},\gamma_{1} \ra\ra_{\ve} & \la\la \gamma_{1},\gamma_{2} \ra\ra_{\ve} \\
                       \la\la \gamma_{2},\gamma_{1} \ra\ra_{\ve} & \la\la \gamma_{2},\gamma_{2} \ra\ra_{\ve} \\
                     \end{array}
                   \right),\, \label{Gr}
\end{equation}
where $\la\la A,B \ra\ra_{\ve}$ denotes the Green's function between operators $A$ and $B$ in energy domain and $G^{a}(\ve)=\left[G^{r}(\ve)\right]^{\dag}$. We find the transmission coefficients for the two setups shown in Fig.\ \ref{Model}, namely models A and B in what follows. These transmission expressions are $\mathcal{T}_{\text{A}}(\ve)$ for the model A and $\mathcal{T}_{\text{B}}(\ve)$ for the model B, and given by \cite{lim2012,flensberg2010}
\begin{equation}
\mathcal{T}_{\text{A}}(\ve)=\frac{4\Gamma^{2}\left(\ve^{2}+\ve_{M}^{2}+4\Gamma^{2}\right)}{(\ve^{2}+4\Gamma^{2})^{2}+\ve_{M}^{2}(\ve_{M}^{2}-2(\ve^{2}-4\Gamma^{2}))}\,, \label{tras1}
\end{equation}
\begin{equation}
\mathcal{T}_{\text{B}}(\ve)=\frac{4\ve^{2}\Gamma^{2}}{\left[(\ve+\ve_{M})(\ve-\ve_{M})\right]^{2}+4\ve^{2}\Gamma^{2}}\,, \label{tras2}
\end{equation}
where $\Gamma$ is the energy-independent coupling strength between the Kitaev chain and the leads for the symmetric case in the wide band limit, where $t_{\alpha,\beta}\equiv t_{0}$ for all non-vanishing cases, and $\Gamma=\pi|t_{0}|^{2}\rho_{0}$, being $\rho_{0}$ the contact density of states.\\

As for thermoelectric quantities, we consider the system in the linear response regime, with a temperature difference $\Delta T$ between the two leads. In this scenario we can write the charge and heat current, $I_{\text{charge}}$ and $I_{\text{heat}}$ respectively, in terms of a potential difference $\Delta V$ as\cite{ziman1960electrons}
\begin{eqnarray}
I_{\text{charge}}&=&-e^{2}L_{0}\Delta V+\frac{e}{T}L_{1}\Delta T\,, \label{Icarga} \\
I_{\text{heat}}&=&e L_{1}\Delta V-\frac{1}{T}L_{2}\Delta T\,, \label{Iheat}
\end{eqnarray}
where $e$ is the electron charge and
\begin{equation}
L_{n}(\mu)=\frac{1}{h}\int\left(-\frac{\partial \bar{f}(\ve,\mu)}{\partial\ve}\right)(\ve-\mu)^{n}\mathcal{T}(\ve)\text{d}\ve\,,
\end{equation}
where $\mu$ and $\bar{f}(\ve,\mu)$ are the Fermi energy and Fermi distribution function respectively, and $h$ the Planck constant. The Seebeck coefficient $S$ (or thermopower) relates the temperature difference $\Delta T$ and the potential difference $\Delta V$ caused when the charge current vanishes,
\begin{equation}
S(\mu)=-\frac{\Delta V}{\Delta T}=-\frac{1}{e\,T}\frac{L_{1}}{L_{0}}\,. \label{S}
\end{equation}
The electrical conductance $\mathcal{G}(\mu)$ and thermal conductance $\kappa(\mu)$, are defined as the ratio between the charge current and the potential difference when $\Delta T$ vanishes for the first, and between the heat current and the temperature gradient when the charge current vanishes for the latter. From Eqs. (\ref{Icarga}) and (\ref{Iheat}), both conductances are given by
\begin{eqnarray}
\mathcal{G}(\mu)&=&-\frac{I_{\text{charge}}}{\Delta V}=e^{2}L_{0}\,, \label{Conductance} \\
\kappa(\mu)&=&-\frac{I_{\text{heat}}}{\Delta T}=\frac{1}{T}\left(L_{2}-\frac{L_{1}^2}{L_{0}}\right)\,. \label{conductanceTh}
\end{eqnarray}
Equation\ (\ref{conductanceTh}) considers only the electronic contribution to the thermal conductance; It assumes that the phononic contribution is negligible in the low-temperature regime (few Kelvin) typical of the systems.\\

In order to quantify the efficiency of our MBS thermoelectric setups, we calculate the dimensionless figure of merit $ZT$,
\begin{equation}
ZT=\frac{S^{2}\mathcal{G}T}{\kappa}\,,
\end{equation}
as function of structure parameters.\\

\section{Results}\label{results}
\subsection{Electrical and Thermal Conductance}\label{subsecA}

In what follows we assume a background temperature of $T=10$ K, well below typical superconductor critical temperatures.\cite{nagamatsu2001superconductivity} We use $\Gamma$ as a useful energy scale and set it to a characteristic experimental value, $\Gamma=10\,\text{meV}$ which leads to $k_{B}T\thicksim10^{-1}\Gamma$, where $k_{B}$ is the Boltzmann constant.

\begin{figure}[tbph]
\centering
\includegraphics[width=0.48\textwidth]{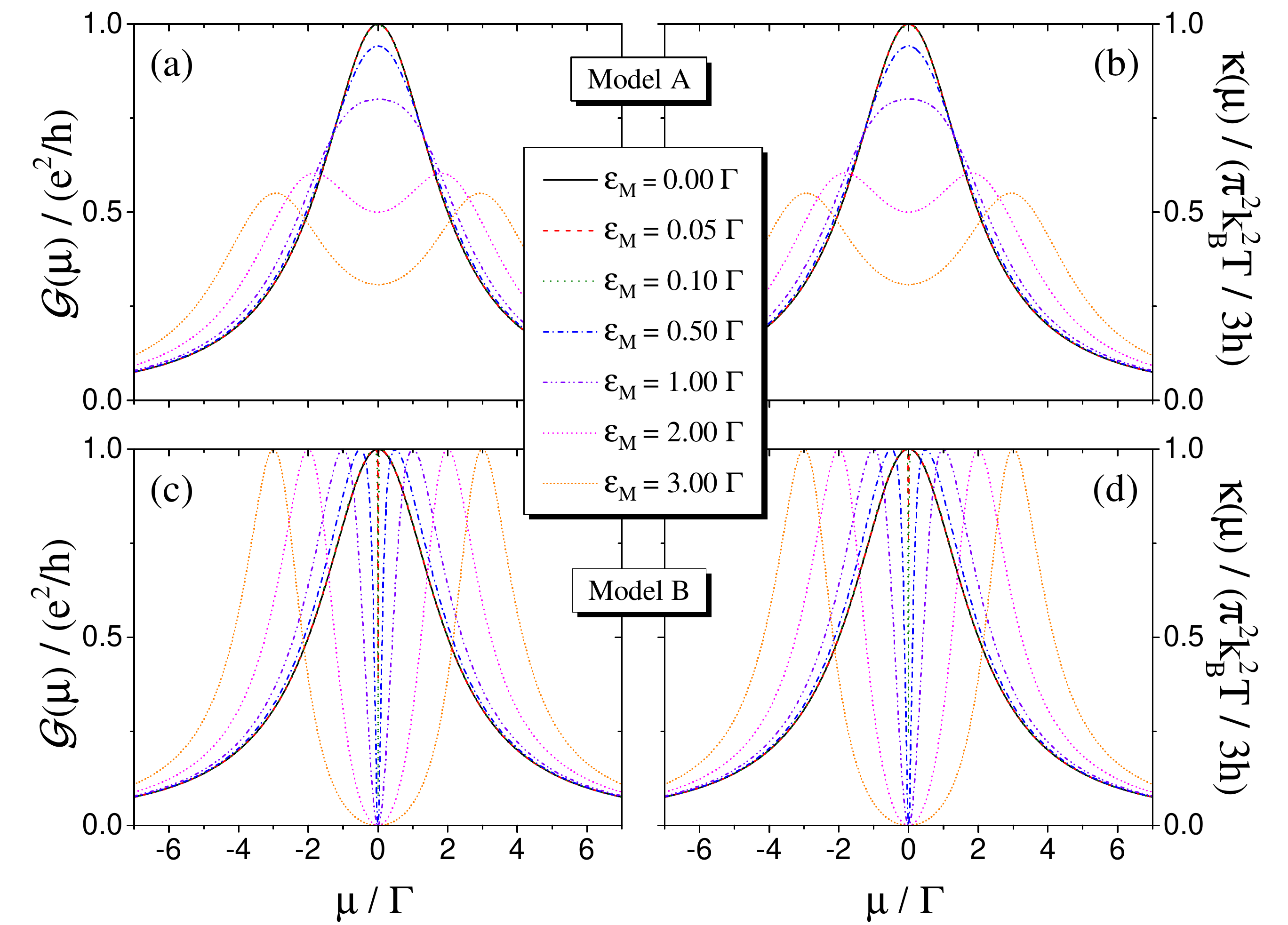}
\caption{(Color online) (a), (c) Electrical and (b), (d) thermal conductance, both as a function of the Fermi energy $\mu$. Upper panels correspond to model A in Fig.\ \ref{Model} and lower panels to Model B respectively.}
\label{ConEandTh}
\end{figure}

For the two setups shown in Fig.\ \ref{Model}, models A and B, Fig.\ \ref{ConEandTh} shows the electrical conductance $\mathcal{G}$ and thermal conductance $\kappa$, in units of $e^{2}/h$ and $\pi^{2}k_{B}^{2}T/3h$, respectively. Figs.\ \ref{ConEandTh}(a) and (b) show $\mathcal{G}$ and $\kappa$ for model A, and Figs.\ \ref{ConEandTh}(c) and (d) show $\mathcal{G}$ and $\kappa$ for model B. In both models, the conductance reaches the maximum value $\mathcal{G}(\mu=0)=e^{2}/h$ when the overlapping parameter $\ve_{M}$ between the two MBS vanishes. The maximun $\mathcal{G}$ occurs whenever the chemical potential of the leads is resonant with the MBSs, as shown in solid black lines. For model A when the $\ve_{M}$ is turned on, such that $0<\ve_{M}\lesssim k_{B}T$, the conductance shows the same behavior, as the central resonance cannot discern the MBS splitting and yields the same maximum magnitude located at $\mu\thickapprox0$. When $\ve_{M}\gtrsim k_{B}T$, there is first a drop in amplitude in the conductance and then, after $\ve_{M}\sim\Gamma$, a clear splitting of the central resonance. For model B, however, the central resonance is split into a central narrow dip at $\mu=0$ and two side peaks at $\pm\ve_{M}$, which reach the same magnitude $\mathcal{G}(\mu=\pm\ve_{M})=e^{2}/h$ in this symmetric coupling case, $\Gamma_{L}=\Gamma_{R}=\Gamma$. The splitting of the central resonance into two side peaks is very evident for $\ve_{M}\gtrsim\Gamma$, with a broad zero near $\mu=0$. Note that both electrical and thermal conductances show the same qualitative behavior, except for a very subtle difference close to the antiresonance located at $\mu=0$, as will be seen later on.\\

\begin{figure}[tbph]
\centering
\includegraphics[width=0.48\textwidth]{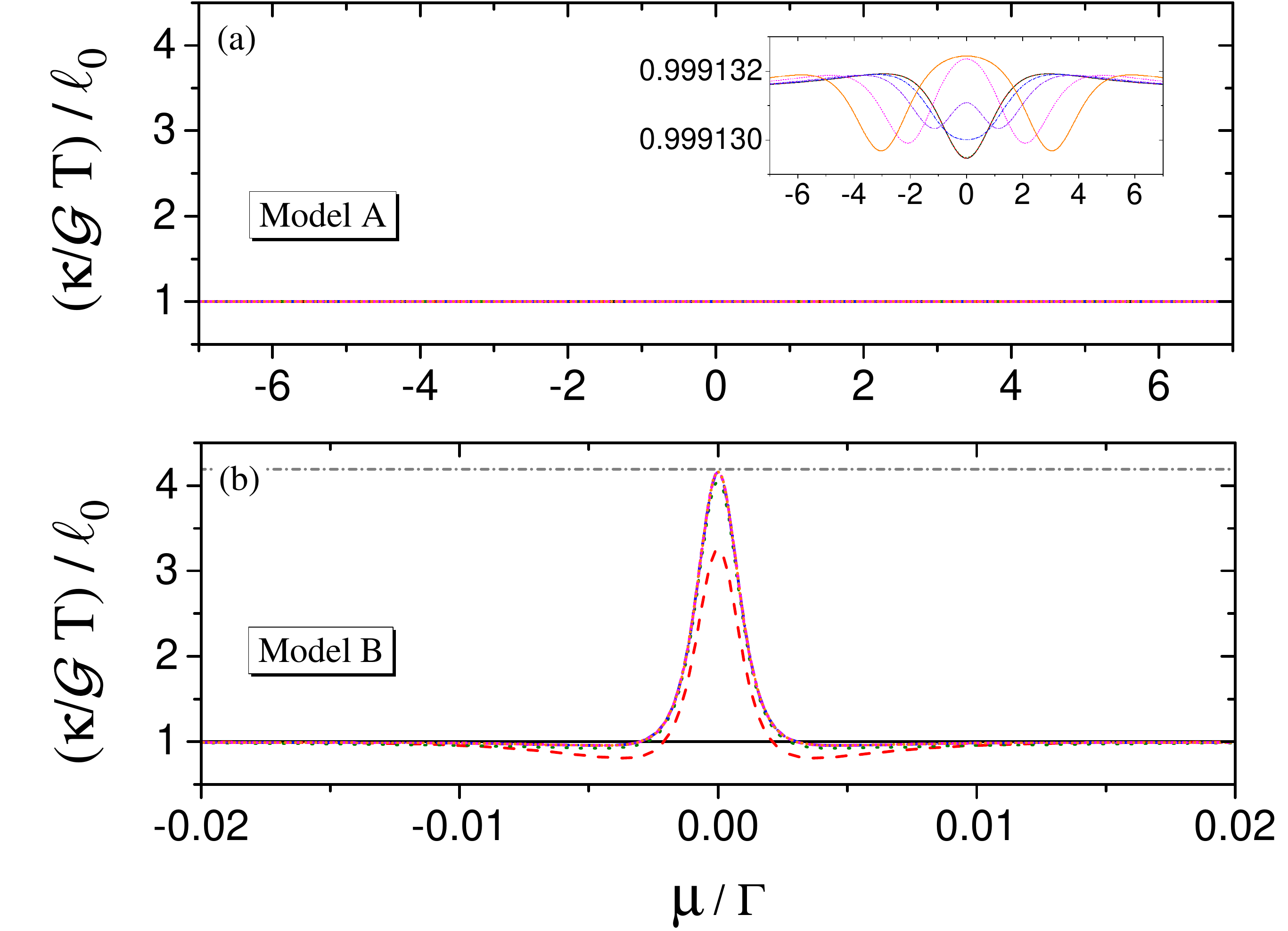}
\caption{(Color online) Wiedemann-Franz law ratio in units of the Lorenz number $\ell_{0}$, for (a) model A, and (b) model B, as in upper and lower panels in Fig.\ \ref{Model}. Horizontal dotted-dashed gray line corresponds to the universal maximum value of $4.19\,\ell_{0}$.\cite{bergfield2009thermoelectric}}
\label{Violation}
\end{figure}

Similar characteristics of the electrical conductance have been discussed in Ref.\ [\onlinecite{wu2012}], as function of the wire length $\mathcal{L}$. By comparison, we can observe that a large (short) $\mathcal{L}$ means weak (strong) MBS overlap $\ve_{M}$ in our model, as one would expect from $\ve_{M}\propto\exp[-\mathcal{L}/\xi]$, where $\xi$ is the superconducting coherence length.\\

\subsection{Wiedemann-Franz law}\label{subsecB}

Let us now explore the fulfilment of the Wiedemann-Franz (WF) law in both geometries by plotting the ratio $\kappa(\mu)/\mathcal{G}(\mu)T$ in Fig.\ \ref{Violation}(a) for model A and in Fig.\ \ref{Violation}(b) for model B, in units of the Lorenz number $\ell_{0}$. For model A we observe a near negligible violation of this law, as the $\kappa/\mathcal{G}T$ ratio is a constant up to the sixth decimal place. Note that $\mathcal{G}(\mu)T>\kappa(\mu)$ is always fulfilled for any $\ve_{M}$, and only the shape of the curves changes for $\ve_{M}\lesssim\Gamma$ and $\ve_{M}\gtrsim\Gamma$, as shown in Fig.\ \ref{Violation}(a). We emphasize that although this deviation from WF is small, it is well within the numerical accuracy of the calculation. For model B, on the other hand, the WF law is fulfilled for $\ve_{M}=0$, but for any $\ve_{M}\neq0$, the violation of the law is observed in a narrow range of $\mu$, rising rapidly to the maximum value $\sim4.19\ell_{0}$ for $\ve_{M}\gtrsim k_{B}T$ at $\mu=0$, as shown in Fig.\ \ref{Violation}(b). This phenomenon is a consequence of the antiresonance in the conductance, similar to those reported in molecules\cite{bergfield2009thermoelectric} and quantum dots\cite{gomez2012}. This drastic violation the Wiedemann-Franz law has not been reported before for systems hosting MBSs.

\begin{figure}[tbph]
\centering
\includegraphics[width=0.48\textwidth]{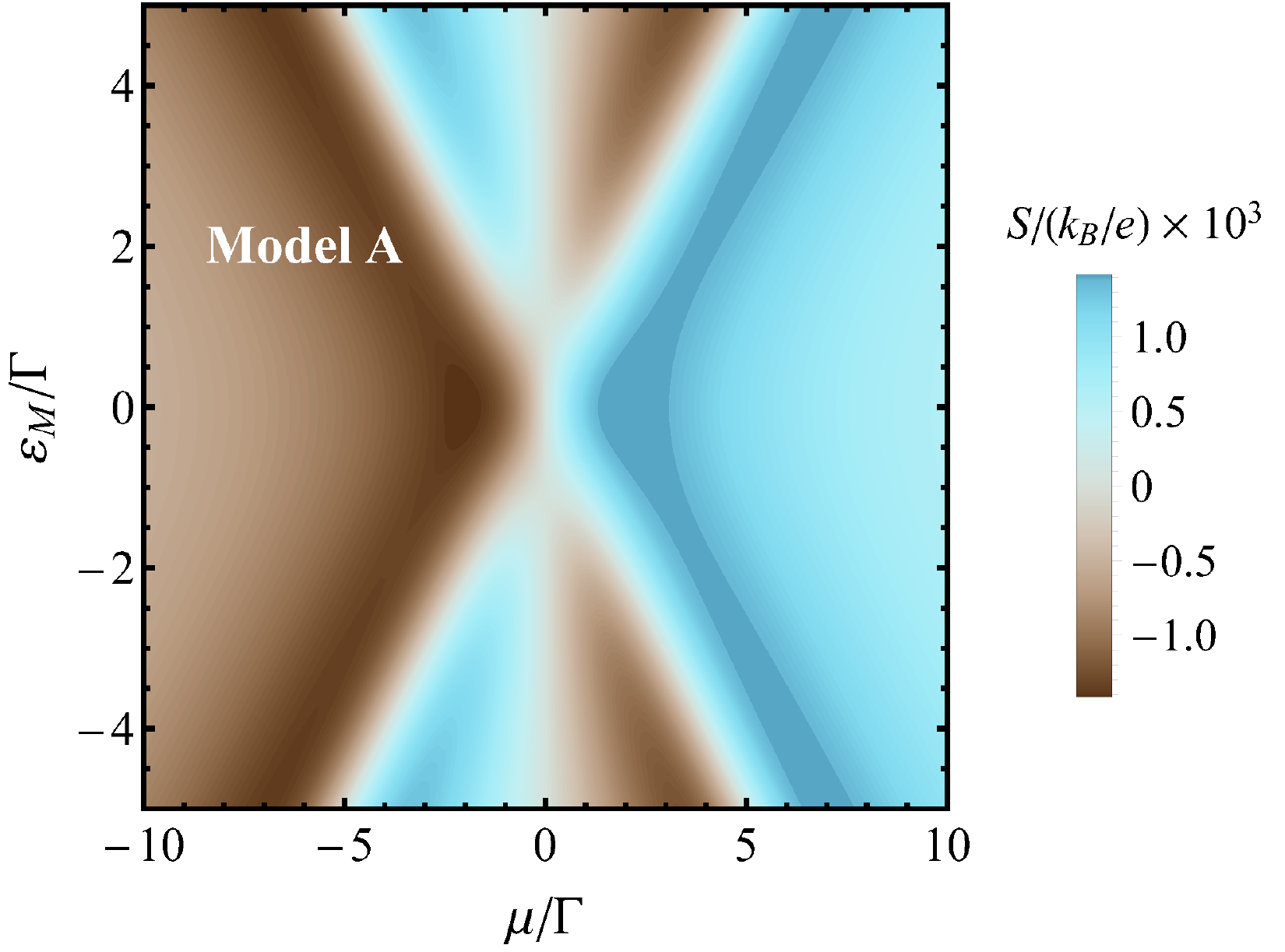}\\
\includegraphics[width=0.48\textwidth]{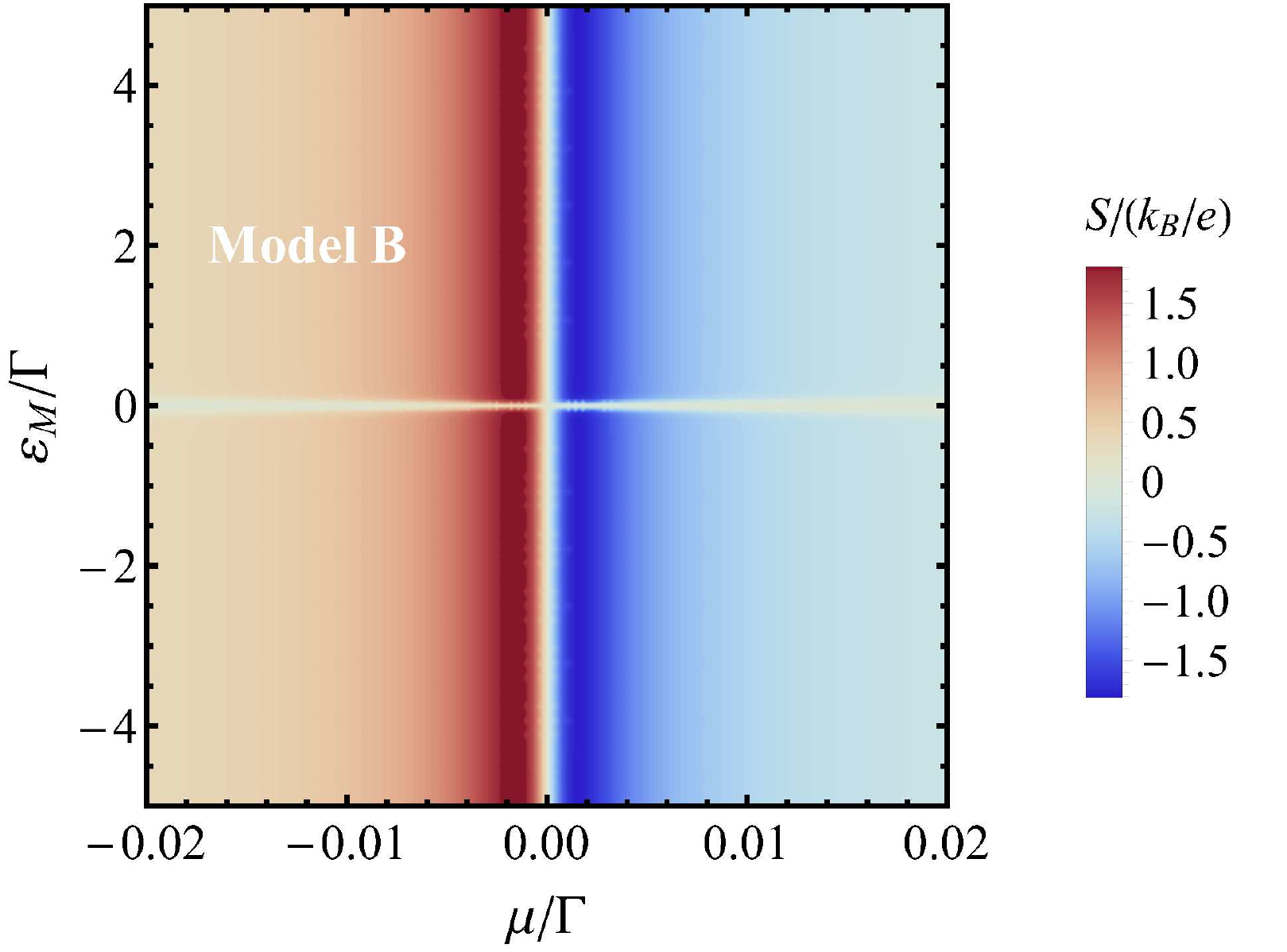}
\caption{(Color online) Seebeck coefficient as a function of $\mu$ and $\ve_{M}$. Upper and lower panels refer to Model A and Model B in Fig.\ \ref{ConEandTh}, respectively. Notice $S$ in Model B can be three orders of magnitude larger than in Model A geometry.}
\label{S}
\end{figure}

\subsection{Thermoelectric efficiency}

\begin{figure}[tbph]
\centering
\includegraphics[width=0.48\textwidth]{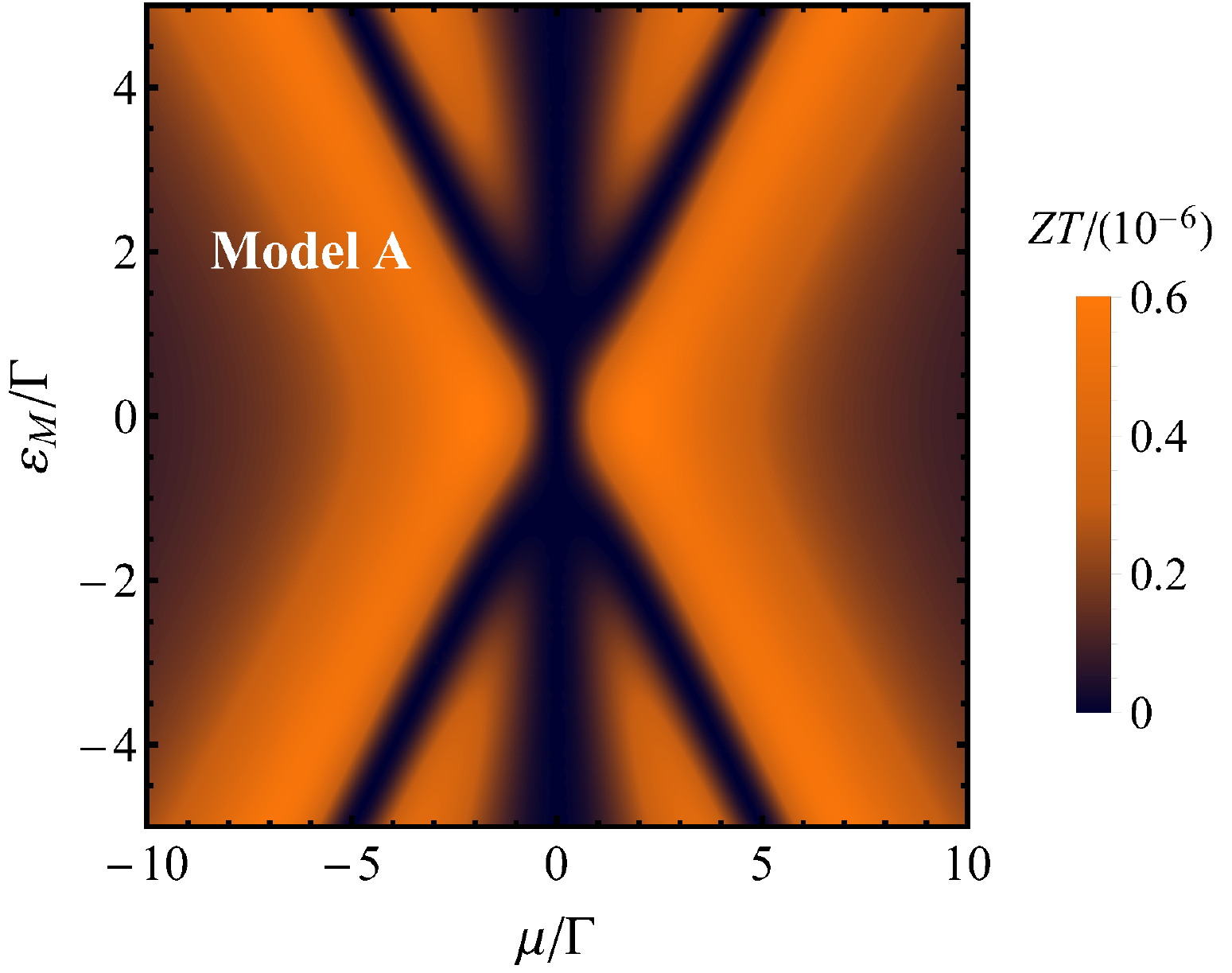}\\
\includegraphics[width=0.48\textwidth]{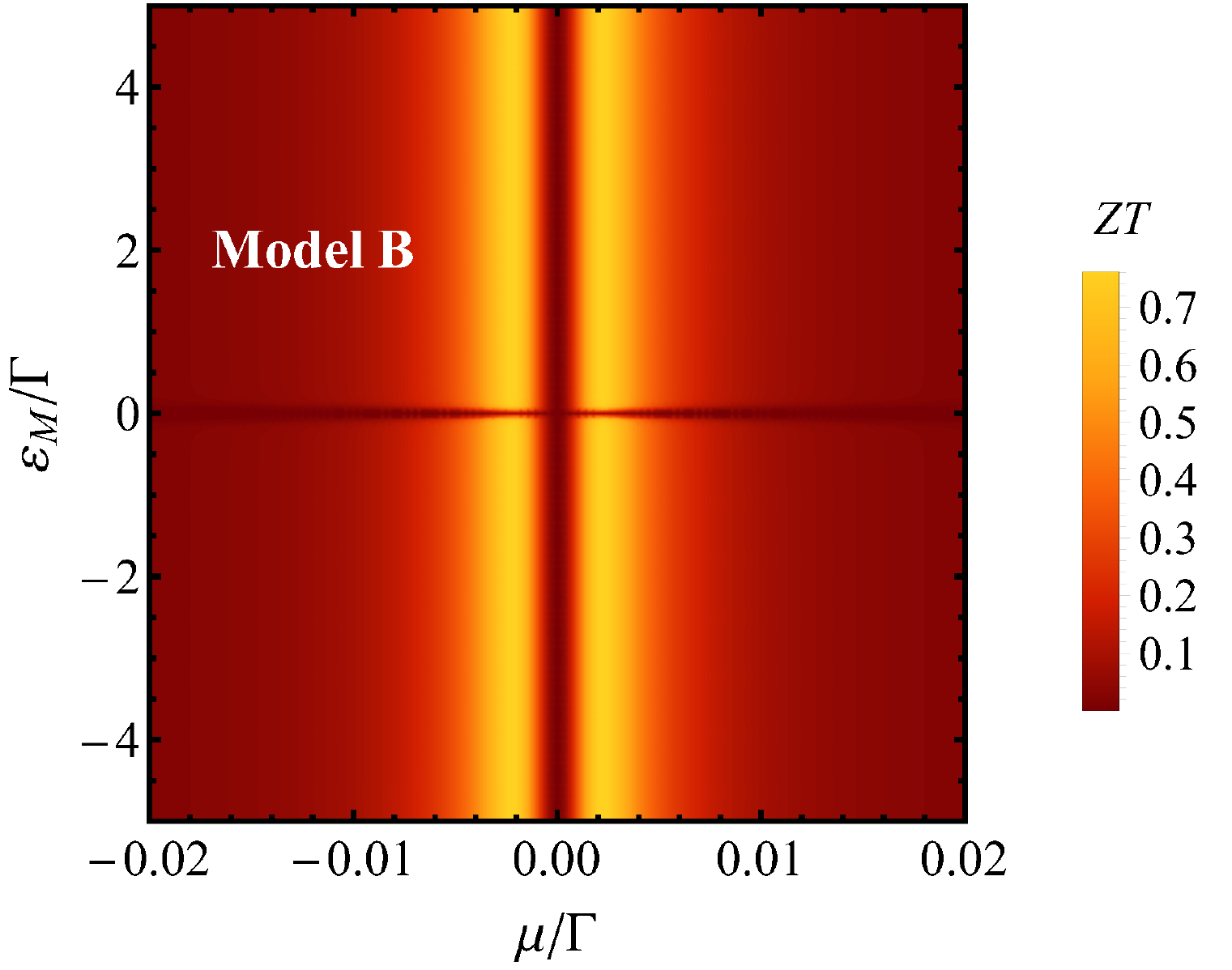}
\caption{(Color online) $ZT$ as a function of $\mu$ and $\ve_{M}$. Upper and lower panels refer to Model A and Model B in Fig.\ \ref{ConEandTh}, respectively. Notice sizeable $ZT$ in Model B over a window $\Delta\mu\sim 0.002\Gamma\sim 20$ $\mu$eV.}
\label{ZT}
\end{figure}

In order to quantify the thermoelectric efficiency of the two geometries, we plot the Seebeck coefficient $(S)$ and figure of merit $(ZT)$ in Figs.\ \ref{S} and \ref{ZT}, respectively. These figures display the vanishing of $S$ and $ZT$ at $\mu=0$, independent of the $\ve_{M}$ values $(S(\mu=0)=ZT(\mu=0)=0)$. The sign of $S$ with respect to $\mu$ depends on the $\ve_{M}$ value for model A, so that for $\mu\leqslant0$ gets $ S(\mu)\leqslant0$ with $|\ve_{M}|\leqslant\Gamma$, but for $|\ve_{M}|>\Gamma$, $S$ changes sign of $\mu$. A similar behavior can be seen for $\mu\geqslant0$. On the other hand, in the lower panel of Fig.\ \ref{S} (model B) the sign of $S$ is essentially independent of $\ve_{M}$, so that $\mu/S(\mu)\leqslant0$ is always obtained, regardless of $\mu$ and $\ve_{M}$. Notice, however, that $S=0$ for $\mu=0$ or/and $\ve_{M}=0$, in sharp contrast to the behavior of model A. Besides, the $\ve_{M}$-gap shown around $\ve_{M}=0$ is proportional to the temperature (not shown). We propose to use the measurement of these features as a signature of the presence of MBSs.

From the upper panel in Fig.\ \ref{ZT}, we can easily see that model A is not thermoelectrically efficient, since $ZT\rightarrow 0\,\,(\sim 10^{-7})$ over the entire parameter domain. In contrast, the lower panel in Fig.\ \ref{ZT}, for model B, shows that the system can be considered thermoelectrically efficient as $ZT$ is near to unity at least in two narrow $\mu$ ranges near zero. It is interesting that the high $ZT$ value is independent of $\ve_{M}$ for $|\ve_{M}|\gtrsim k_{B}T$.

\section{Conclusions}\label{secconclusions}

We have studied the thermoelectric transport through a nanowire hosting MBSs, when a temperature gradient is applied. We find that when only one end of the nanowire is connected to normal metal leads sustaining a thermal gradient, the $ZT$ figure of merit approaches 1 for small deviations of the chemical potencial away from zero. Although experiments to explore this phenomenon would require control of $\Delta\mu\sim 20$ $\mu$eV, they would provide unique signatures of MBS in these systems.

\section{Acknowledgments}\label{secacknowledgment}
J. P. R.-A. is grateful for the hospitality of Ohio University and the funding of scholarship CONICYT-Chile No 21141034. P. A. O. acknowledges support from FONDECYT grant No. 1140571 and CONICYT ACT 1204. S. E. U. and O. \'A.-O. acknowledge support from NSF Grant No. DMR 1508325.

\noindent

\bibliographystyle{apsrev4-1}
\bibliography{biblio}
\nocite{*}

\end{document}